\documentclass[twocolumn]{aastex631}

\usepackage{textcomp, gensymb}

\shorttitle{No Metastable Helium in LTT 9779b}
\shortauthors{Vissapragada et al.}

\begin{document}

\correspondingauthor{Shreyas~Vissapragada}
\email{shreyas.vissapragada@cfa.harvard.edu}

\title{A High-Resolution Non-Detection of Escaping Helium In The Ultra-Hot Neptune LTT 9779b: Evidence for Weakened Evaporation}

\author[0000-0003-2527-1475]{Shreyas~Vissapragada}
\altaffiliation{51 Pegasi b Fellow}
\affiliation{Center for Astrophysics $\vert$ Harvard \& Smithsonian, 60 Garden Street, Cambridge, MA 02138, USA}

\author[0000-0003-0473-6931]{Patrick McCreery}
\affiliation{Department of Physics and Astronomy, Johns Hopkins University, 3400 N Charles St, Baltimore, MD 21218, USA}

\author[0000-0002-2248-3838]{Leonardo A. Dos Santos}
\affiliation{Space Telescope Science Institute, 3700 San Martin Drive, Baltimore, MD 21218, USA}

\author[0000-0001-9513-1449]{N\'{e}stor Espinoza}
\affiliation{Space Telescope Science Institute, 3700 San Martin Drive, Baltimore, MD 21218, USA}

\author{Andrew McWilliam}
\affiliation{The Observatories of the Carnegie Institution for Science, 813 Santa Barbara Street, Pasadena, CA 91101}

\author{Noriyuki Matsunaga} 
\affiliation{Department of Astronomy, School of Science, The University of Tokyo, 7-3-1 Hongo, Bunkyo-ku, Tokyo 113-0033, Japan}
\affiliation{Laboratory of Infrared High-resolution spectroscopy (LiH), Koyama Astronomical Observatory, Kyoto Sangyo University, Motoyama, Kamigamo, Kita-ku, Kyoto, 603-8555, Japan}


\author[0000-0002-4489-3168]{J\'{e}a Adams Redai}
\affiliation{Center for Astrophysics $\vert$ Harvard \& Smithsonian, 60 Garden Street, Cambridge, MA 02138, USA}

\author[0000-0003-1369-8551]{Patrick Behr}
\affiliation{Department of Astrophysical \& Planetary Sciences, University of Colorado Boulder, Boulder, CO, 80309, USA}
\affiliation{Laboratory for Atmospheric and Space Physics, University of Colorado Boulder, Boulder, CO, 80303, USA}

\author[0000-0002-1002-3674]{Kevin France}
\affiliation{Department of Astrophysical \& Planetary Sciences, University of Colorado Boulder, Boulder, CO, 80309, USA}
\affiliation{Laboratory for Atmospheric and Space Physics, University of Colorado Boulder, Boulder, CO, 80303, USA}

\author[0000-0002-6505-3395]{Satoshi Hamano} 
\affiliation{National Astronomical Observatory of Japan, 2-21-1 Osawa, Mitaka, Tokyo 181-8588, Japan}

\author{Charlie Hull}
\affiliation{The Observatories of the Carnegie Institution for Science, 813 Santa Barbara Street, Pasadena, CA 91101}

\author[0000-0003-2380-8582]{Yuji Ikeda} 
\affiliation{Photocoding, 460-102 Iwakura-Nakamachi, Sakyo-ku, Kyoto 606-0025, Japan}
\affiliation{Laboratory of Infrared High-resolution spectroscopy (LiH), Koyama Astronomical Observatory, Kyoto Sangyo University, Motoyama, Kamigamo, Kita-ku, Kyoto, 603-8555, Japan}

\author{Haruki Katoh} 
\affiliation{Laboratory of Infrared High-resolution spectroscopy (LiH), Koyama Astronomical Observatory, Kyoto Sangyo University, Motoyama, Kamigamo, Kita-ku, Kyoto, 603-8555, Japan}

\author{Hideyo Kawakita} 
\affiliation{Laboratory of Infrared High-resolution spectroscopy (LiH), Koyama Astronomical Observatory, Kyoto Sangyo University, Motoyama, Kamigamo, Kita-ku, Kyoto, 603-8555, Japan}
\affiliation{Department of Astrophysics and Atmospheric Sciences, Faculty of Science, Kyoto Sangyo University, Motoyama, Kamigamo, Kita-ku, Kyoto 603-8555, Japan}

\author[0000-0003-3204-8183]{Mercedes L\'{o}pez-Morales}
\affiliation{Center for Astrophysics $\vert$ Harvard \& Smithsonian, 60 Garden Street, Cambridge, MA 02138, USA}

\author[0000-0003-3455-8814]{Kevin N. Ortiz Ceballos}
\affiliation{Center for Astrophysics $\vert$ Harvard \& Smithsonian, 60 Garden Street, Cambridge, MA 02138, USA}

\author{Shogo Otsubo} 
\affiliation{Laboratory of Infrared High-resolution spectroscopy (LiH), Koyama Astronomical Observatory, Kyoto Sangyo University, Motoyama, Kamigamo, Kita-ku, Kyoto, 603-8555, Japan}

\author{Yuki Sarugaku} 
\affiliation{Laboratory of Infrared High-resolution spectroscopy (LiH), Koyama Astronomical Observatory, Kyoto Sangyo University, Motoyama, Kamigamo, Kita-ku, Kyoto, 603-8555, Japan}

\author{Tomomi Takeuchi} 
\affiliation{Laboratory of Infrared High-resolution spectroscopy (LiH), Koyama Astronomical Observatory, Kyoto Sangyo University, Motoyama, Kamigamo, Kita-ku, Kyoto, 603-8555, Japan}

\begin{abstract}
The recent discovery of ``ultra-hot'' ($P < 1$~day) Neptunes has come as a surprise: some of these planets have managed to retain gaseous envelopes despite being close enough to their host stars to trigger strong photoevaporation and/or Roche lobe overflow. Here, we investigate atmospheric escape in LTT 9779b, an ultra-hot Neptune with a volatile-rich envelope. We observed two transits of this planet using the newly-commissioned WINERED spectrograph ($R\sim68,000$) on the 6.5~m Clay/Magellan II Telescope, aiming to detect an extended upper atmosphere in the He 10830~\AA~triplet. We found no detectable planetary absorption: in a 0.75~\AA~passband centered on the triplet, we set a 2$\sigma$ upper limit of 0.12\% ($\delta R_p/H < 14$) and a 3$\sigma$ upper limit of 0.20\% ($\delta R_p/H < 22$). Using a H/He isothermal Parker wind model, we found corresponding 95\% and 99.7\% upper limits on the planetary mass-loss rate of $\dot{M} < 10^{10.03}$~g~s$^{-1}$ and $\dot{M} < 10^{11.11}$~g~s$^{-1}$ respectively, smaller than predicted by outflow models even considering the weak stellar XUV emission. The low evaporation rate is plausibly explained by a metal-rich envelope, which would decrease the atmospheric scale height and increase the cooling rate of the outflow. This hypothesis is imminently testable: if metals commonly weaken planetary outflows, then we expect that \textit{JWST} will find high atmospheric metallicities for small planets that have evaded detection in He 10830~\AA.
\end{abstract}

\section{Introduction} \label{sec:intro}

Ultra-hot Neptunes are an enigmatic new part of exoplanet population. Prior to the launch of the \textit{Transiting Exoplanet Survey Satellite} (\textit{TESS}) mission, there were no Neptune-mass planets known with orbital periods of less than a day, despite our ability to detect them. The dearth of ultra-hot Neptunes has been variously attributed to tidal disruption during high-eccentricity migration, photoevaporation, and/or Roche-lobe overflow \citep{Kurokawa2014, Matsakos2016, Mazeh2016, OwenLai2018, Ionov2018, Koskinen2022, Thorngren2023}. Over the past few years, \textit{TESS} has revealed that the Neptune desert is not empty \citep{Jenkins2020, Armstrong2020, Persson2022, Naponiello2023, Osborn2023}. Studying the properties of desert-dwelling planets may help us understand the physical mechanisms that clear out the Neptune desert and drive the evolution of short-period exoplanets. 

LTT 9779b was the first ultra-hot Neptune to be unveiled by \textit{TESS}. At $M_p = 29.32^{+0.78}_{-0.81}M_\Earth$ and $R_p = 4.72\pm0.23R_\Earth$, the planet has a density of just $\rho_p = 1.54\pm0.12$~g~cm$^{-3}$ \citep{Jenkins2020}. It therefore appears to be volatile-rich despite remarkable proximity ($P = 0.79$~d, $a = 0.017$~au) to a G7V host star. LTT 9779b's orbital separation is only $1.5\times$ the Roche limit for an incompressible fluid $a_\mathrm{Roche} = 2.44R_\star(\rho_\star/\rho_p)^{1/3}$, so it may be tidally distorted as well. For planets at LTT 9779b's orbital distance, \citet{Koskinen2022} predict the 1~bar substellar radius to be $\sim5\%$ larger than the polar radius for hot Jupiter and hot Neptune analogs \citep[see also e.g.][]{Li2010, Rappaport2013, Delrez2016}.

This planet is a compelling target for atmospheric spectroscopy, especially for studies of atmospheric erosion. Assuming the envelope composition is dominated by H$_2$/He ($\mu = 2.3$~amu), the scale height of LTT 9779b's atmosphere at $R_p$ is 550~km, similar to many hot Jupiters that have been studied with \textit{HST} and \textit{JWST}. Also, the planet has a restricted Jeans parameter $\Lambda = GM_pm_\mathrm{H} / (k_\mathrm{B}T_\mathrm{eq}R_p) \approx 27$, a gravitational potential $\Phi = GM_p/R_p \approx 4\times10^{12}$~erg~g$^{-1}$, and a Roche lobe filling factor $R_p/R_\mathrm{Roche}\approx 50\%$ \citep{Eggleton1983}. Photoevaporation and/or Roche lobe overflow should be vigorous for a planet so close to its host star with such low $\rho_p$, $\Lambda$, $\Phi$, and $R_\mathrm{Roche}$ \citep{Salz2016, Kubyshkina2018, Caldiroli2022, Edwards2023}. In fact, escaping atmospheres have been detected for many Jupiter- and Neptune-sized planets with similarly small $\rho_p$, $\Lambda$, $\Phi$, and/or $R_\mathrm{Roche}$ \citep{DosSantos2023}.

LTT 9779b is therefore an excellent target for transmission spectroscopy in the metastable He 10830~\AA~triplet, which probes atmospheric escape \citep{Oklopcic2018, Spake2018}. Planets orbiting stars cooler than the Sun \citep{Oklopcic2019} are now routinely studied in He 10830~\AA, allowing for population-level constraints on planetary mass-loss rates \citep{Vissapragada2022:helium, Allart2023, Guilluy2023, DosSantos2023}. Models suggest that planets orbiting late G stars like LTT 9779 should be favorable targets for helium outflow observations \citep{Oklopcic2019, Wang2021, Biassoni2023}, but there have been relatively few He 10830~\AA~studies for planets orbiting G-type stars \citep{Allart2023, Bennett2023, Guilluy2023}. Recently, \citet{Edwards2023} attempted to constrain helium absorption in the upper atmosphere of LTT 9779b with the \textit{HST} Wide-Field Camera 3 (WFC3) in the G102 grism, but resolving power was ultimately a limiting factor. Using an updated version of the model by \citet{Allan2019}, these authors predicted 0.3\% absorption in the core of the helium triplet (assuming a 90/10 H/He number ratio), too small to be detected by \textit{HST} WFC3/G102 at $R\sim100$. Ground-based high-resolution spectroscopy can be used to fully resolve the line core, especially in a system as bright as LTT 9779 ($J = 8.4$), but most instruments sensitive to He~10830~\AA~are located in the northern hemisphere. At a declination of $-37\degree$, LTT 9779b has thus far eluded high-resolution near-infrared transit observations.

Recently, the Warm INfrared Echelle spectrograph to REalize Decent high-resolution spectroscopy \citep[WINERED; ][]{Ikeda2022} was installed on the 6.5~m Clay/Magellan II Telescope to help fill the need for near-infrared high-resolution spectrographs in the southern hemisphere. In this work, we use WINERED to constrain helium absorption in LTT 9779b at high resolving power ($R\sim68,000$). This is the first time that the WINERED spectrograph has been used to search for helium in an exoplanet atmosphere. In Section~\ref{sec:obs}, we describe our WINERED observations of LTT 9779b as well as our data reduction methodology. We model our He~10830~\AA~non-detection in Section~\ref{sec:mod} to constrain the planetary mass-loss rate. Finally, we discuss our findings in Section~\ref{sec:disc} and conclude in Section~\ref{sec:conc}.

\section{Observations and Data Reduction} \label{sec:obs}

We observed two transits of LTT 9779b on UT 2023 June 8 and June 12 using the WINERED spectrograph \citep{Ikeda2016, Ikeda2022} on the 6.5~m Clay/Magellan II Telescope at Las Campanas Observatories. Data were obtained in HIRES-Y mode \citep{Otsubo2016} with the 100~$\mu$m slit (slit width 9$\arcsec$), corresponding to a resolving power of $R\sim68,000$. We took 180~s exposures in an ABBA nod pattern (5$\arcsec$ throw between positions) to correct the OH airglow lines, although for a few strong OH lines we observed some remaining residuals which were later masked. On the first night, we took 64 exposures between UT 06:48 and UT 10:44, covering phases from 0.87 to 0.08 and corresponding to a starting airmass of 1.9 and a final airmass of 1.0. Seeing was excellent throughout the night at around $0\farcs5$. On the second night, conditions were somewhat worse, with the seeing hovering around $1\farcs0$. We took 76 exposures between UT 06:00 and UT 10:22, covering phases from 0.88 to 0.11 and corresponding to a starting airmass of 2.3 and a final airmass of 1.0. 

\begin{figure*}
    \centering
    \includegraphics[width=\textwidth]{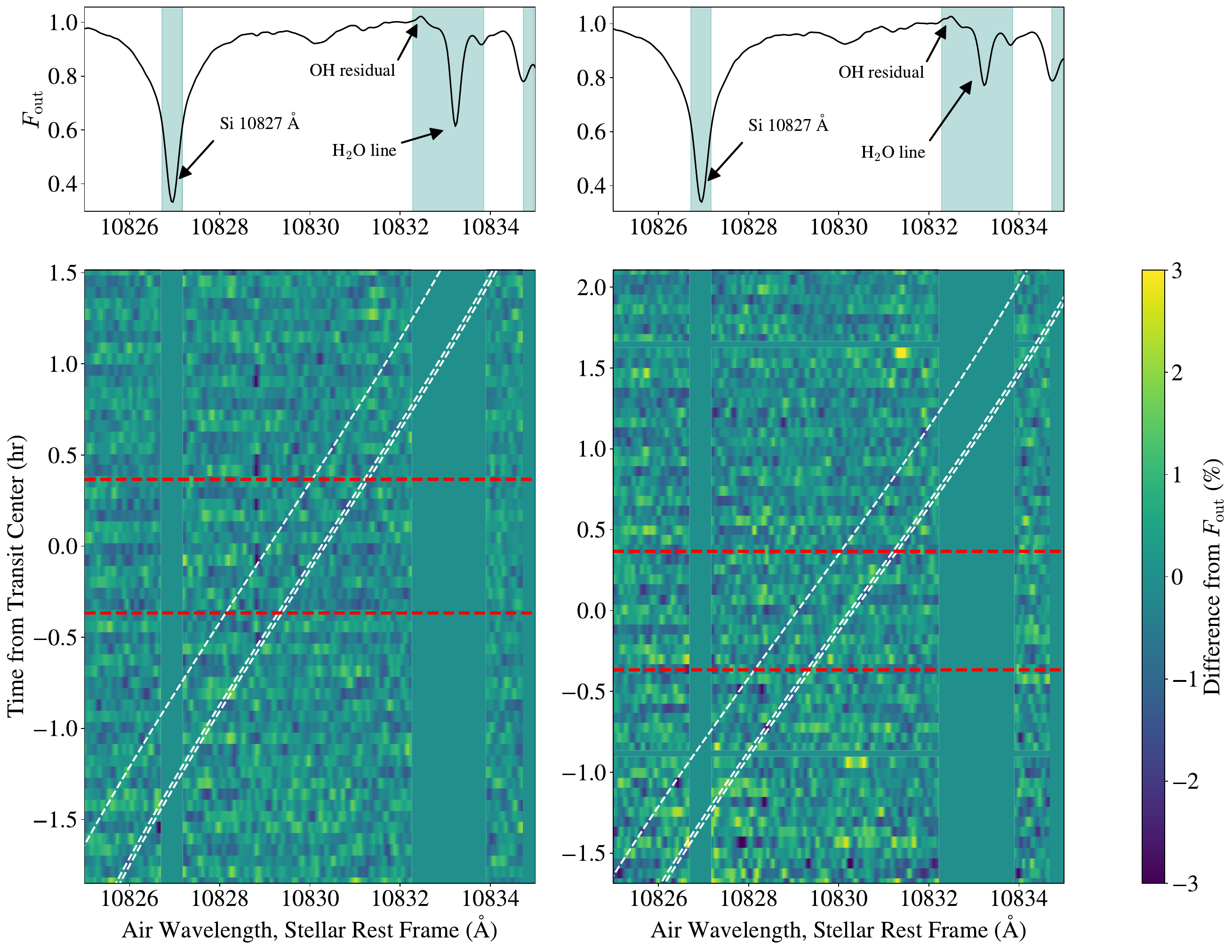}
    \caption{Timeseries spectroscopy of LTT 9779 in the stellar rest frame on the first (left) and second (right) nights of data collection. The colors indicate percentage difference from the average out-of-transit spectrum, shown in the upper panel for each night. Red dashed lines indicate the beginning and end of the transit on each night, and the diagonal white lines indicate the positions of the helium triplet lines following the planetary rest frame. The masks for the tellurics and the Si 10827~\AA~line core are shown with the green vertical bars and annotated in the upper panels, and the two low S/N frames on the second night are indicated by the green horizontal bars.}
    \label{fig:stellarrestframe}
\end{figure*}

We reduced the spectra using the WINERED Automatic Reduction Pipeline\footnote{https://github.com/SatoshiHamano/WARP} (WARP ver3.8.12; Hamano et al. in preparation). After flat-fielding and subtracting sky emission and scattered light, WARP performs aperture extraction and wavelength calibration using a ThAr lamp spectrum. We used the fluxes and wavelengths produced by the WARP pipeline for order 163, which includes the helium triplet. On the first night, we obtained a median S/N of 108 and on the second night we obtained a median S/N of 89. Two frames from the second night were removed due to low $\mathrm{S/N}< 20$ exposures. 

We also noticed persistence in our data: when moving from position A to B, a residual trace of about 10\% remained at position A, and vice versa. Internal WINERED tests have shown the persistence drops to negligible levels after 30~min, which is longer than the time between subsequent exposures in both of our transit timeseries, and typically longer than the setup time for a given target (meaning that persistence from previous observations of bright stars may affect the first few images). We therefore discarded the first 30~min of data (8 exposures) on each night before performing our analysis, which allowed the residual image from previous exposures to dissipate and the differential persistence at the A and B positions to stabilize. For future helium observations on WINERED, we advocate including a 30~min on-target overhead period to accommodate the persistence effects. 

We proceeded to normalize and align the spectra for timeseries analysis. First, we continuum-normalized the spectra by fitting a fourth-degree Chebyshev polynomial to the line-free regions. We tested using lower-degree polynomials, but found that this increased correlated noise in the final transmission spectrum. Next, we refined the wavelength solution by comparing our spectra to a line-by-line telluric transmission spectrum from HITRAN \citep[calculated with the \textsf{hapi} package;][]{Kochanov2016, Gordon2022}. After accounting for the instrumental broadening with a Gaussian kernel, we cross-correlated the telluric template with the average spectrum on each night. The determined shifts were then corrected to ensure that all data were aligned in the telluric rest frame.

After verifying the wavelength solution, we masked the telluric features for our analysis, following e.g. \citet{Orell-Miquel2022} and \citet{Spake2022}. Imperfect telluric correction may introduce extra correlated noise to the final transmission spectrum, which can make it more challenging to estimate a meaningful upper limit in the case of a non-detection. This is especially important when considering the WINERED detector persistence. Telluric absorption and emission may change rapidly even from exposure to exposure, and because the detector has some memory of previous exposures, the observed telluric features for any given exposure are a weighted average over time. We therefore adopt the conservative approach of masking the tellurics near the helium triplet, including water absorption lines and regions with residuals from OH sky subtraction.

Finally, we shifted the data into the stellar rest frame. We used \textsf{astropy} to first shift spectra into the Solar System barycenter (SSB) frame, and then shifted into the stellar rest frame using the systemic radial velocity of $-10.72$~km~s$^{-1}$ \citep{GaiaDR3} relative to SSB. Before continuing the analysis, we masked the cores of the deep Mg 10811~\AA, Si 10827~\AA, and Si 10844~\AA~lines, as we noticed weak ($\lesssim1\%$) time-correlated variations at these wavelengths. Such residuals are commonly seen and masked in helium observations \citep[e.g.][]{Guilluy2023, Zhang2023} and likely originate from the inadequacy of typical spectral extraction approaches for sharp lines in high-SNR data \citep{Zhang2021}.

\begin{figure*}
    \centering
    \includegraphics[width=.99\textwidth]{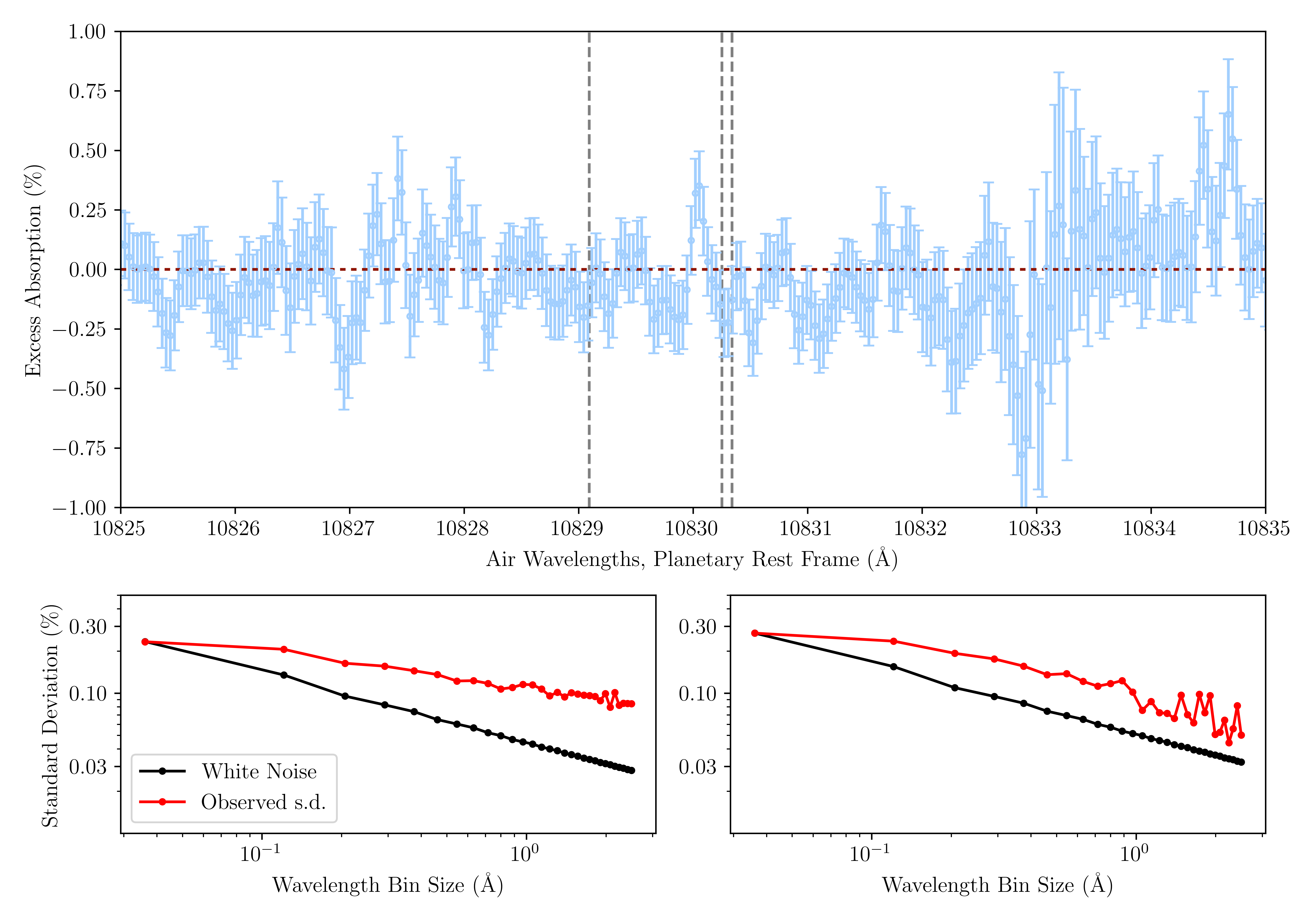} 
    \caption{Transmission spectrum of LTT 9779b averaged over both nights of observations (top) and Allan deviation plots for night 1 (bottom left) and night 2 (bottom right). In the transmission spectrum, excess absorption in the planetary rest frame is shown with the blue points, and the dashed red curve indicates no excess absorption. The helium triplet wavelength positions are marked with vertical dashed lines, and the increased uncertainty near 10833~\AA~is due to the masking of telluric features. For visual clarity, the transmission spectrum does not include the $\beta$ factors that was used to inflate the photon-noise uncertainties during the \textsf{p-winds} fit. In the bottom panels, the Allan deviation plots show the observed rms of the data as a function of bin size (red) compared to the expectations for white noise (black), indicating a substantial correlated noise component for both nights.}
    \label{fig:transmissionspectra}
\end{figure*}

With these features masked, we constructed a combined out-of-transit stellar spectrum for each night, shown in Figure~\ref{fig:stellarrestframe}. Out-of-transit spectra were identified using the linear ephemerides from \citet{Edwards2023}, which are precise to better than 30~s for our transits. Each spectroscopic timeseries was then normalized by its combined out-of-transit spectrum to produce the final timeseries spectra. Our final timeseries spectra are shown in Figure~\ref{fig:stellarrestframe}. No excess helium absorption during transit is evident. To obtain the final planetary transmission spectrum, we shifted these timeseries into the planetary rest frame and took the mean of all the in-transit spectra. The results are shown in Figure~\ref{fig:transmissionspectra}. The transmission spectrum is noticeably less precise in wavelength bins that are affected by masked tellurics near 10833~\AA. Again, no absorption is detected near the 10830~\AA~triplet in the planetary rest frame. 

To set an upper limit on the planetary absorption, we first calculate the mean absorption in a 0.75~\AA~bin centered on the helium triplet ($-0.04\%$), and we take $\sigma$ to be the rms scatter of the transmission spectrum binned to 0.75~\AA~($0.08\%$). Estimating $\sigma$ in this way accounts for correlated noise \citep{Allart2023, Guilluy2023}. Our resulting $2\sigma$ and $3\sigma$ upper limits in a 0.75~\AA~bin centered on the helium triplet are 0.12\% and 0.20\%, respectively. These correspond to 14 and 22 lower-atmospheric scale heights respectively ($\delta R_\mathrm{p}/H < 14$ at $2\sigma$ and $\delta R_\mathrm{p}/H < 22$ at 3$\sigma$), calculated as $H = \frac{k_\mathrm{B}T_\mathrm{eq}}{\mu g}$ with $\mu = 2.3$~amu. Our reporting is slightly different than \citet{Allart2023} and \citet{Guilluy2023}, who report 3$\sigma$ for upper limits without the contribution of the mean absorption, and calculate $H$ assuming $\mu = 1.3$~amu. Under their assumptions our upper limits are 0.24\% and $\delta R_\mathrm{p}/H < 14$ (the larger $\delta R_p$ is balanced out by the larger $H$, resulting in nearly the same $\delta R_\mathrm{p}/H$ as our $2\sigma$ upper limit).

\section{Modeling} \label{sec:mod}

We used the open source, one-dimensional, isothermal H+He Parker wind code \textsf{p-winds}\footnote{\footnotesize{Version 1.4.4 (doi:10.5281/zenodo.7814782). Publicly available in \url{https://github.com/ladsantos/p-winds}.}} \citep{DSantos2022} to study the implications of our non-detection for the upper atmosphere of LTT 9779b. As in the original isothermal Parker wind helium models from \citet{Oklopcic2018} and \citet{Lampon2020}, \textsf{p-winds} computes the expected transmission spectrum for the helium triplet in an isothermal wind given an atmospheric escape rate $\dot{M}$ and an upper atmosphere temperature $T$ along with an estimate for the top-of-atmosphere XUV spectrum, the H/He number ratio (assumed to be 90/10), well-known planetary parameters like the mass and radius, and the instrumental line spread function (modeled as a Gaussian with a FWHM of 0.16~\AA). Using our observed transmission spectrum in the planetary rest frame, we retrieve posterior probability distributions on $\dot{M}$ and $T$.  In the case of our non-detection, we are able to place upper limits on the mass-loss rate. 

Our retrieval relies on an estimate for the high-energy spectral energy distribution (SED) of LTT 9779, which is a slowly-rotating ($v\sin i = 1.06\pm0.37$ km~s$^{-1}$) and relatively inactive ($\log R_{HK}' = -5.10 \pm 0.04$) star \citep{Jenkins2020}. We used the MUSCLES Extension SED for TOI-193/LTT 9779 from \citet{Behr2023}. The SED includes a scaled spectrum of the quiet Sun for the X-ray ($5-100$~\AA) and FUV ($1170-2180$~\AA), as the star was not detected in a 22.9~ks \textit{Chandra}/ACIS-S observation. The stellar EUV ($100-1170$~\AA) emission was estimated using the scaling relations from \citet{Linsky2014}. Some FUV emission lines were reconstructed as Gaussian profiles based on the upper limit fluxes from \textit{HST} STIS/G140L data. Lyman-$\alpha$ was reconstructed using the MCMC methods described in \citet{Youngblood2022}. The SED includes STIS/G230L data from ($2180-3110$~\AA) and STIS/G430L data from ($3110-5700$~\AA). A BT-Settl CIFIST stellar model was used for wavelengths 5700~\AA~and greater.

Before we retrieved posterior probability distributions on our parameters of interest, we needed to account for additional systematic uncertainties arising from correlated noise in the planetary transmission spectrum. Long-wavelength structure is clearly seen in Figure~\ref{fig:transmissionspectra}. To account for this additional source of uncertainty, we first constructed an Allan deviation plot for each night's transmission spectrum, calculating the true standard deviation of the data at varying wavelength bin sizes ($\sigma_r$) and comparing this to the theoretical expectation for pure photon noise ($\sigma_w$) \citep{Pont2006, Carter2009, Allart2023}. We then inflated the uncertainties on each transmission spectrum by the factor \begin{equation}
\beta = \sqrt{1 + \left(\frac{\sigma_r}{\sigma_w}\right)^2} \mathrm{,}
\end{equation} where $\sigma_r$ and $\sigma_w$ were estimated at a bin size of 0.75~\AA. We found $\beta = 1.8$ for the first night's transmission spectrum and $\beta = 1.7$ for the second night's transmission spectrum. An approach using Gaussian processes to better account for correlated noise is currently being developed, which will improve mass-loss constraints (McCreery et al., in preparation).

\begin{figure}
    \centering
    \includegraphics[width=0.5\textwidth]{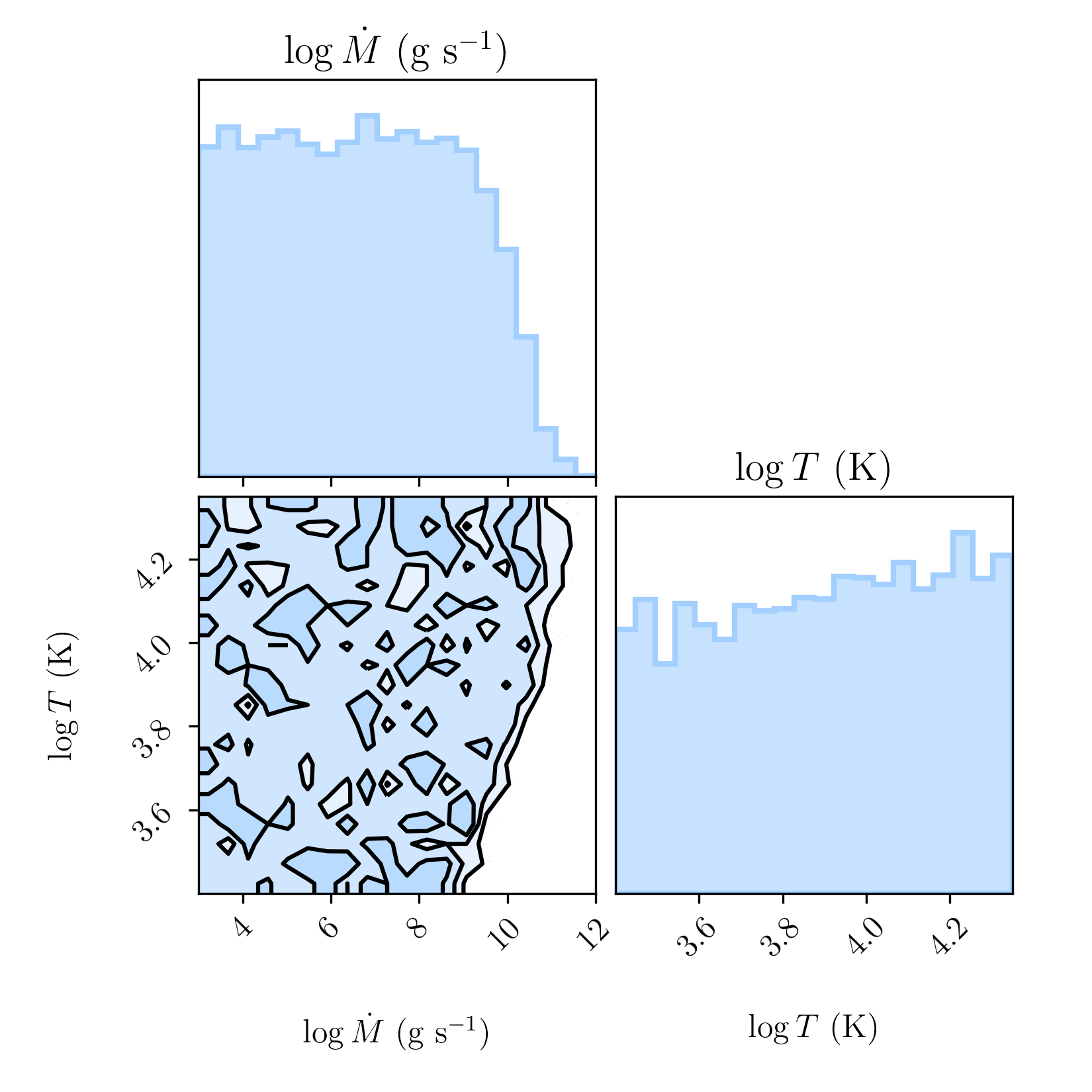} 
    \caption{Posterior probability distributions on our sampled parameters. The plotted contours are the $1\sigma$, $2\sigma$, and $3\sigma$ levels for the 2D distribution.}
    \label{fig:posterior}
\end{figure}

We used nested sampling \citep{Skilling2004, Skilling2006} to constrain the posterior probability distributions for $\dot{M}$ and $T$. Using the \textsf{dynesty}\footnote{\footnotesize{Version 2.1.2, doi: 10.5281/zenodo.3348367}} package \citep{Speagle2020}, we sample across a prior volume defined by $\mathcal{U}(3, 12.5)$ for $\log\dot{M}$ and $\mathcal{U}(3.35, 4.35)$ for $\log T$. We perform dynamic nested sampling \citep{Higson2019} with 500 initial live points, using the \textsf{'multi'} method (i.e. multiple ellipsoids) for bounding and the  \textsf{'unif'} method (i.e. uniformly sampling within the bounds) for sampling. Convergence is achieved when the estimated change in evidence from further sampling does not exceed $\Delta\log\mathcal{Z} > 0.5$. The posterior distributions that we obtained are shown in Figure~\ref{fig:posterior}. 

We estimate an 95th-percentile upper limit of $\dot{M} < 10^{10.03}$~g~s$^{-1}$ and a 99.7th-percentile upper limit of $\dot{M} < 10^{11.11}$~g~s$^{-1}$. The latter corresponds to a mass-loss timescale of $M_p/\dot{M} \gtrsim 40$~Gyr, i.e. negligible for the planet's evolution. The outflow temperature is weakly constrained by our non-detection, similarly to e.g. \citet{Spake2022}. Outflows close to our upper limit on $\dot{M}$ are permitted only at high temperatures $T\gtrsim10,000$~K, but at lower temperatures the constraint on $\dot{M}$ is more stringent. We re-ran the analysis using \textsf{emcee} \citep{Foreman-Mackey2013}, and verified that the joint posterior probability distribution on $\dot{M}$ and $T_0$ was robust to our choice of sampling methodology.

\section{Discussion} \label{sec:disc}
The non-detection of helium in LTT 9779b's atmosphere is somewhat puzzling. It is unlikely to be due to helium ionization: while some short-period planets have reduced neutral triplet helium populations due to strong photoionization \citep{Oklopcic2019, Biassoni2023}, the top-of-atmosphere XUV flux for LTT 9779b is actually quite modest. Using the \citet{Behr2023} SED, we found that LTT 9779b had a top-of-atmosphere XUV flux (integrated between 5~\AA~and 504~\AA, corresponding to helium-ionizing wavelengths) of 4800~erg~s$^{-1}$~cm$^{-2}$: similar to that of HD 189733b, half that of WASP-69b, and only about 3$\times$ larger than that of HAT-P-11b \citep{Allart2023}. We therefore do not expect LTT 9779b's upper atmosphere to be substantially more ionized than these three planets, which all exhibit strong helium absorption signatures. It is more likely that either LTT 9779b's mass-loss rate or helium abundance is low.

To determine if a low mass-loss rate is expected for this planet, we compared our upper limits ($\dot{M} < 10^{10.03}$~g~s$^{-1}$ at 95\% confidence, $\dot{M} < 10^{11.11}$~g~s$^{-1}$ at 99.7\% confidence) to predicted $\dot{M}$ values from various photoevaporation models. These models require the integrated XUV flux shortward of 912~\AA~(i.e. photons capable of photoionizing hydrogen and/or helium), which we estimated to be $F_\mathrm{XUV} = 6500$~erg~cm$^{-2}$~s$^{-1}$ from the MUSCLES spectrum. Using the ATES model grid from \citet{Caldiroli2022}, we calculated a predicted mass-loss rate of $\dot{M} = 7.3\times10^{10}$~g~s$^{-1}$. This is in good agreement with \citet{Edwards2023}, who also used the MUSCLES spectrum from \citet{Behr2023} and predicted a mass-loss rate of $\dot{M} \approx 10^{11}$~g~s$^{-1}$ with a detailed outflow model based on \citet{Allan2019}. These mass-loss rates are ruled out by our observations at the 95\% level. Recently, \citet{FernandezFernandez2024} have obtained even deeper constraints on the X-ray luminosity of LTT 9779 with \textit{XMM-Newton}. These authors estimated $\dot{M} = 1.5\times10^{11}$~g~s$^{-1}$ using the model from \citet{Kubyshkina2018}, and $\dot{M} = 6\times10^{10}$~g~s$^{-1}$ using an energy-limited model. Both of these mass-loss rates are also ruled out by our observations at the 95\% level (and the former at the 99.7\% level). The discrepancy between the helium non-detection and the outflow models is more severe when we consider Roche lobe overflow, which at these distances should be greatly enhancing the outflow rate over expectations from photoevaporation alone \citep{Koskinen2022}. We conclude that the outflow rate of LTT 9779b appears to be smaller than expected, even considering the low XUV luminosity of the host star.

A number of additional factors could be playing a role in the non-detection of metastable helium. For example, outflows can be confined by the stellar wind \citep{Spake2021, Wang2021, MacLeod2022} or planetary magnetic field \citep{Fossati2023, Schreyer2023}. These scenarios lead to different velocity structures of the helium triplet, but in the case of a non-detection we lack the velocity information necessary to assess these possibilities. Another assumption of our retrieval was a 90/10 H/He number ratio. If LTT 9779b's atmosphere is substantially depleted in helium, our inferred total mass-loss rate would increase, bringing the inference into better agreement with photoevaporation models. Helium depletion has been invoked to help explain weak detections and non-detections in other systems, including the non-detection in WASP-80b \citep{Fossati2023, Lampon2023}. 

Our modeling has also assumed a metal-free H/He atmosphere, which a number of recent investigations call into question for LTT 9779b. Phase curve and thermal emission observations from \textit{Spitzer}, \textit{TESS}, and \textit{CHEOPS} suggest that the planet is extremely reflective, with geometric albedo $A_g = 0.80^{+0.10}_{-0.17}$ \citep{Crossfield2020, Dragomir2020, Hoyer2023}. \citet{Hoyer2023} estimated that a highly supersolar ($>400\times$ solar) metallicity is required for such reflective clouds to survive on the planetary dayside. These results are in some tension with \citet{Edwards2023}, who estimate the metallicity to be sub-solar based on their \textit{HST} WFC3 transmission spectrum. The \textit{JWST} NIRISS/SOSS phase curves taken in Cycle 1 (GTO 1201) and planned for Cycle 2 (GO 3231) will help to resolve the discrepancy.

A high atmospheric metallicity would naturally explain the low inferred mass-loss rate for LTT 9779b. As atmospheric metallicity increases past $100\times$ solar, the scale height of the upper atmosphere is substantially decreased and thermospheric cooling becomes dominated by metals, decreasing both the overall mass-loss rate and the He 10830~\AA~signal \citep{Owen2012, Lopez2017, Owen2018, Ito2021, Nakayama2022, Zhang2022}. Weakened evaporation due to metals may also be responsible for the non-detections of He~10830~\AA~ in other sub-Jovian systems, including GJ 1214b \citep{Kasper2020, Spake2022}, which  \textit{JWST} has revealed to be quite metal-rich \citep{Gao2023, Kempton2023}. For LTT 9779b, a metal-rich envelope may be the final product of rapid atmospheric loss, which is more likely from Roche lobe overflow or partial tidal disruption than photoevaporation \citep{Guillochon2011, Koskinen2022, Vissapragada2022:helium, Osborn2023}. Ultra-hot Neptunes may have started their lives as gas giants before catastrophic envelope loss ensued and deeper, metal-enriched gas was exposed, making the planet resilient to further photoevaporation \citep{Jenkins2020, Hoyer2023}. This might also explain why some hot Neptunes survive well into the post-main sequence evolution of their host stars \citep{Grunblatt2023}.

\section{Conclusion} \label{sec:conc}
We observed two transits of LTT 9779b using the WINERED spectrograph on the 6.5~m Clay/Magellan II telescope. Our aim was to search for excess absorption in the metastable helium triplet during transit, which would have indicated an extended upper atmosphere. We did not detect any excess absorption: in a 0.75~\AA~passband centered on the helium triplet, we set upper limits of $<0.12\%$ ($\delta R_\mathrm{p}/H < 14$) and $<0.20\%$ ($\delta R_\mathrm{p}/H < 22$) at 2$\sigma$ and 3$\sigma$ respectively. Using an isothermal Parker wind model, we found the planetary mass-loss rate to be $<10^{10.03}$~g~s$^{-1}$ and $<10^{11.11}$~g~s$^{-1}$ at the 95th and 99.7th percentiles respectively. Our model accounted for the relatively low XUV luminosity of LTT 9779, which certainly contributes to the non-detection \citep{Behr2023, FernandezFernandez2024}. Even so, our inferred mass-loss rate is at least a factor of a few smaller than predicted by photoevaporation models that assume a pure H/He composition for the planetary envelope \citep{Kubyshkina2018, Caldiroli2022, Edwards2023, FernandezFernandez2024}. 

We suggest that the non-detection of helium may be due in part to a high envelope metallicity, which has been suggested independently based on \textit{Spitzer}, \textit{TESS}, and \textit{CHEOPS} phase curve observations \citep{Crossfield2020, Hoyer2023}. Metals can efficiently cool planetary outflows, leading to smaller mass-loss rates than the pure H/He models predict. Additionally, the higher mean atomic weight of the thermosphere would depress the atmospheric scale height, making features like He 10830~\AA~more challenging to detect. If atmospheric metals do indeed drive weakened evaporation in planets like LTT 9779b, we expect that \textit{JWST} will reveal a metal-rich atmosphere for this planet, and probably also for other small planets that have evaded detection in He 10830~\AA~\citep{Kasper2020, Carleo2021, Spake2022, Allart2023, Guilluy2023}. Conversely, if the atmospheric metallicity of LTT 9779b is not found to be significantly super-solar, the helium non-detection could be due to a low helium abundance \citep{Lampon2023}, a strong stellar wind \citep{MacLeod2022}, or the planetary magnetic field \citep{Schreyer2023}. 

LTT 9779b remains one of the best desert-dwelling planets for transmission spectroscopy, and it should be prioritized for further study given the non-detection of helium escape. The \textit{JWST} Cycle 1 NIRISS/SOSS phase curve of LTT 9779b (GTO 1201) will cover the He 10830~\AA~triplet wavelengths at high precision \citep{Fu2022, DosSantos2023}. Jointly analyzing these data with the WINERED transmission spectrum will provide even stronger constraints on the helium absorption. Additionally, the H$\alpha$ line has emerged as a powerful probe of atmospheric escape in ultra-hot Jupiters \citep[e.g.][]{Jensen2012, Yan2018, Wyttenbach2020, Huang2023}, making it an interesting direction forward for ultra-hot Neptunes. A detection of escape in H$\alpha$ or Lyman-$\alpha$ might suggest LTT 9779b is helium-depleted rather than metal-rich \citep[e.g.][]{Yan2022, Lampon2023}. However, Lyman-$\alpha$ spectroscopy is challenging at $d = 81$~pc due to both the diminished stellar flux and the strong interstellar absorption. A better option in the ultraviolet may be spectroscopy of metal lines in the near-ultraviolet with \textit{HST} \citep[e.g.][]{Sing2019, Linssen2023} or a future dedicated small-sat platform \citep{France2023, Sreejith2023}. If the atmospheres of ultra-hot Neptunes are indeed metal-rich, then near-UV observations may be the best way to study their upper atmospheres.

\begin{acknowledgements}
This paper is based on WINERED data gathered with the 6.5 meter Clay/Magellan II Telescope located at Las Campanas Observatory, Chile. We thank the staff at Las Campanas Observatory for their efforts to ensure a successful first science run for WINERED. In particular we thank Jorge Araya for assistance with telescope operations, and we also thank Morgan Saidel, Heather Knutson, and Yuri Beletsky for helpful conversations. 

WINERED was developed by the University of Tokyo and the Laboratory of Infrared High-resolution Spectroscopy, Kyoto Sangyo University, under the financial support of KAKENHI (Nos. 16684001, 20340042, and 21840052) and the MEXT Supported Program for the Strategic Research Foundation at Private Universities (Nos. S0801061 and S1411028). The observing run in 2023 June was partly supported by KAKENHI (grant No 18H01248) and JSPS Bilateral Program Number JPJSBP120239909. A. McWilliam thanks and acknowledges receipt of a Carnegie Venture Grant, kindly provided by the Carnegie Institution for Science, in order to purchase equipment required to adapt, install and support WINERED on the Magellan Clay telescope.
\end{acknowledgements}

\facilities{Magellan:Clay (WINERED spectrograph), ADS, Exoplanet Archive}
\software{\textsf{numpy} \citep{numpy}, \textsf{matplotlib} \citep{Hunter2007}, \textsf{scipy} \citep{scipy}, \textsf{astropy} \citep{exoplanet:astropy13, exoplanet:astropy18}, \textsf{spectres} \citep{Carnall2017}, \textsf{hapi} \citep{Kochanov2016}, \textsf{p-winds} \citep{DSantos2022}, \textsf{dynesty} \citep{Speagle2020}, \textsf{emcee} \citep{Foreman-Mackey2013}}
\clearpage
\bibliography{references}

\begin{thebibliography}{}
\expandafter\ifx\csname natexlab\endcsname\relax\def\natexlab#1{#1}\fi
\providecommand{\url}[1]{\href{#1}{#1}}
\providecommand{\dodoi}[1]{doi:~\href{http://doi.org/#1}{\nolinkurl{#1}}}
\providecommand{\doeprint}[1]{\href{http://ascl.net/#1}{\nolinkurl{http://ascl.net/#1}}}
\providecommand{\doarXiv}[1]{\href{https://arxiv.org/abs/#1}{\nolinkurl{https://arxiv.org/abs/#1}}}

\bibitem[{{Allan} \& {Vidotto}(2019)}]{Allan2019}
{Allan}, A., \& {Vidotto}, A.~A. 2019, \mnras, 490, 3760,
  \dodoi{10.1093/mnras/stz2842}

\bibitem[{{Allart} {et~al.}(2023){Allart}, {Lem{\'e}e-Joliecoeur}, {Jaziri},
  {Lafreni{\`e}re}, {Artigau}, {Cook}, {Darveau-Bernier}, {Dang}, {Cadieux},
  {Boucher}, {Bourrier}, {Deibert}, {Pelletier}, {Radica}, {Benneke},
  {Carmona}, {Cloutier}, {Cowan}, {Delfosse}, {Donati}, {Doyon}, {Figueira},
  {Forveille}, {Fouqu{\'e}}, {Gaidos}, {Gu}, {H{\'e}brard}, {Kiefer},
  {K{\'o}sp{\'a}l}, {Jayawardhana}, {Martioli}, {Dos Santos}, {Turner}, \&
  {Vidotto}}]{Allart2023}
{Allart}, R., {Lem{\'e}e-Joliecoeur}, P.~B., {Jaziri}, A.~Y., {et~al.} 2023,
  arXiv e-prints, arXiv:2307.05580, \dodoi{10.48550/arXiv.2307.05580}

\bibitem[{{Armstrong} {et~al.}(2020){Armstrong}, {Lopez}, {Adibekyan}, {Booth},
  {Bryant}, {Collins}, {Deleuil}, {Emsenhuber}, {Huang}, {King}, {Lillo-Box},
  {Lissauer}, {Matthews}, {Mousis}, {Nielsen}, {Osborn}, {Otegi}, {Santos},
  {Sousa}, {Stassun}, {Veras}, {Ziegler}, {Acton}, {Almenara}, {Anderson},
  {Barrado}, {Barros}, {Bayliss}, {Belardi}, {Bouchy}, {Brice{\~n}o}, {Brogi},
  {Brown}, {Burleigh}, {Casewell}, {Chaushev}, {Ciardi}, {Collins},
  {Col{\'o}n}, {Cooke}, {Crossfield}, {D{\'\i}az}, {Delgado Mena}, {Demangeon},
  {Dorn}, {Dumusque}, {Eigm{\"u}ller}, {Fausnaugh}, {Figueira}, {Gan},
  {Gandhi}, {Gill}, {Gonzales}, {Goad}, {G{\"u}nther}, {Helled}, {Hojjatpanah},
  {Howell}, {Jackman}, {Jenkins}, {Jenkins}, {Jensen}, {Kennedy}, {Latham},
  {Law}, {Lendl}, {Lozovsky}, {Mann}, {Moyano}, {McCormac}, {Meru},
  {Mordasini}, {Osborn}, {Pollacco}, {Queloz}, {Raynard}, {Ricker}, {Rowden},
  {Santerne}, {Schlieder}, {Seager}, {Sha}, {Tan}, {Tilbrook}, {Ting}, {Udry},
  {Vanderspek}, {Watson}, {West}, {Wilson}, {Winn}, {Wheatley}, {Villasenor},
  {Vines}, \& {Zhan}}]{Armstrong2020}
{Armstrong}, D.~J., {Lopez}, T.~A., {Adibekyan}, V., {et~al.} 2020, \nat, 583,
  39, \dodoi{10.1038/s41586-020-2421-7}

\bibitem[{{Astropy Collaboration} {et~al.}(2013){Astropy Collaboration},
  {Robitaille}, {Tollerud}, {Greenfield}, {Droettboom}, {Bray}, {Aldcroft},
  {Davis}, {Ginsburg}, {Price-Whelan}, {Kerzendorf}, {Conley}, {Crighton},
  {Barbary}, {Muna}, {Ferguson}, {Grollier}, {Parikh}, {Nair}, {Unther},
  {Deil}, {Woillez}, {Conseil}, {Kramer}, {Turner}, {Singer}, {Fox}, {Weaver},
  {Zabalza}, {Edwards}, {Azalee Bostroem}, {Burke}, {Casey}, {Crawford},
  {Dencheva}, {Ely}, {Jenness}, {Labrie}, {Lim}, {Pierfederici}, {Pontzen},
  {Ptak}, {Refsdal}, {Servillat}, \& {Streicher}}]{exoplanet:astropy13}
{Astropy Collaboration}, {Robitaille}, T.~P., {Tollerud}, E.~J., {et~al.} 2013,
  \aap, 558, A33, \dodoi{10.1051/0004-6361/201322068}

\bibitem[{{Astropy Collaboration} {et~al.}(2018){Astropy Collaboration},
  {Price-Whelan}, {Sip{\H o}cz}, {G{\"u}nther}, {Lim}, {Crawford}, {Conseil},
  {Shupe}, {Craig}, {Dencheva}, {Ginsburg}, {VanderPlas}, {Bradley},
  {P{\'e}rez-Su{\'a}rez}, {de Val-Borro}, {Aldcroft}, {Cruz}, {Robitaille},
  {Tollerud}, {Ardelean}, {Babej}, {Bach}, {Bachetti}, {Bakanov}, {Bamford},
  {Barentsen}, {Barmby}, {Baumbach}, {Berry}, {Biscani}, {Boquien}, {Bostroem},
  {Bouma}, {Brammer}, {Bray}, {Breytenbach}, {Buddelmeijer}, {Burke},
  {Calderone}, {Cano Rodr{\'{\i}}guez}, {Cara}, {Cardoso}, {Cheedella},
  {Copin}, {Corrales}, {Crichton}, {D'Avella}, {Deil}, {Depagne}, {Dietrich},
  {Donath}, {Droettboom}, {Earl}, {Erben}, {Fabbro}, {Ferreira}, {Finethy},
  {Fox}, {Garrison}, {Gibbons}, {Goldstein}, {Gommers}, {Greco}, {Greenfield},
  {Groener}, {Grollier}, {Hagen}, {Hirst}, {Homeier}, {Horton}, {Hosseinzadeh},
  {Hu}, {Hunkeler}, {Ivezi{\'c}}, {Jain}, {Jenness}, {Kanarek}, {Kendrew},
  {Kern}, {Kerzendorf}, {Khvalko}, {King}, {Kirkby}, {Kulkarni}, {Kumar},
  {Lee}, {Lenz}, {Littlefair}, {Ma}, {Macleod}, {Mastropietro}, {McCully},
  {Montagnac}, {Morris}, {Mueller}, {Mumford}, {Muna}, {Murphy}, {Nelson},
  {Nguyen}, {Ninan}, {N{\"o}the}, {Ogaz}, {Oh}, {Parejko}, {Parley}, {Pascual},
  {Patil}, {Patil}, {Plunkett}, {Prochaska}, {Rastogi}, {Reddy Janga},
  {Sabater}, {Sakurikar}, {Seifert}, {Sherbert}, {Sherwood-Taylor}, {Shih},
  {Sick}, {Silbiger}, {Singanamalla}, {Singer}, {Sladen}, {Sooley},
  {Sornarajah}, {Streicher}, {Teuben}, {Thomas}, {Tremblay}, {Turner},
  {Terr{\'o}n}, {van Kerkwijk}, {de la Vega}, {Watkins}, {Weaver}, {Whitmore},
  {Woillez}, {Zabalza}, \& {Astropy Contributors}}]{exoplanet:astropy18}
{Astropy Collaboration}, {Price-Whelan}, A.~M., {Sip{\H o}cz}, B.~M., {et~al.}
  2018, \aj, 156, 123, \dodoi{10.3847/1538-3881/aabc4f}

\bibitem[{{Behr} {et~al.}(2023){Behr}, {France}, {Brown}, {Duvvuri}, {Bean},
  {Berta-Thompson}, {Froning}, {Miguel}, {Pineda}, {Wilson}, \&
  {Youngblood}}]{Behr2023}
{Behr}, P.~R., {France}, K., {Brown}, A., {et~al.} 2023, \aj, 166, 35,
  \dodoi{10.3847/1538-3881/acdb70}

\bibitem[{{Bennett} {et~al.}(2023){Bennett}, {Redfield}, {Oklop{\v{c}}i{\'c}},
  {Carleo}, {Ninan}, \& {Endl}}]{Bennett2023}
{Bennett}, K.~A., {Redfield}, S., {Oklop{\v{c}}i{\'c}}, A., {et~al.} 2023, \aj,
  165, 264, \dodoi{10.3847/1538-3881/acd34b}

\bibitem[{{Biassoni} {et~al.}(2023){Biassoni}, {Caldiroli}, {Gallo}, {Haardt},
  {Spinelli}, \& {Borsa}}]{Biassoni2023}
{Biassoni}, F., {Caldiroli}, A., {Gallo}, E., {et~al.} 2023, arXiv e-prints,
  arXiv:2310.13052, \dodoi{10.48550/arXiv.2310.13052}

\bibitem[{{Caldiroli} {et~al.}(2022){Caldiroli}, {Haardt}, {Gallo}, {Spinelli},
  {Malsky}, \& {Rauscher}}]{Caldiroli2022}
{Caldiroli}, A., {Haardt}, F., {Gallo}, E., {et~al.} 2022, \aap, 663, A122,
  \dodoi{10.1051/0004-6361/202142763}

\bibitem[{{Carleo} {et~al.}(2021){Carleo}, {Youngblood}, {Redfield}, {Casasayas
  Barris}, {Ayres}, {Vannier}, {Fossati}, {Palle}, {Livingston}, {Lanza},
  {Niraula}, {Alvarado-G{\'o}mez}, {Chen}, {Gandolfi}, {Guenther}, {Linsky},
  {Nagel}, {Narita}, {Nortmann}, {Shkolnik}, \& {Stangret}}]{Carleo2021}
{Carleo}, I., {Youngblood}, A., {Redfield}, S., {et~al.} 2021, \aj, 161, 136,
  \dodoi{10.3847/1538-3881/abdb2f}

\bibitem[{{Carnall}(2017)}]{Carnall2017}
{Carnall}, A.~C. 2017, arXiv e-prints, arXiv:1705.05165,
  \dodoi{10.48550/arXiv.1705.05165}

\bibitem[{{Carter} \& {Winn}(2009)}]{Carter2009}
{Carter}, J.~A., \& {Winn}, J.~N. 2009, \apj, 704, 51,
  \dodoi{10.1088/0004-637X/704/1/51}

\bibitem[{{Crossfield} {et~al.}(2020){Crossfield}, {Dragomir}, {Cowan},
  {Daylan}, {Wong}, {Kataria}, {Deming}, {Kreidberg}, {Mikal-Evans}, {Gorjian},
  {Jenkins}, {Benneke}, {Collins}, {Burke}, {Henze}, {McDermott}, {Mireles},
  {Watanabe}, {Wohler}, {Ricker}, {Vanderspek}, {Seager}, \&
  {Jenkins}}]{Crossfield2020}
{Crossfield}, I. J.~M., {Dragomir}, D., {Cowan}, N.~B., {et~al.} 2020, \apjl,
  903, L7, \dodoi{10.3847/2041-8213/abbc71}

\bibitem[{{Delrez} {et~al.}(2016){Delrez}, {Santerne}, {Almenara}, {Anderson},
  {Collier-Cameron}, {D{\'\i}az}, {Gillon}, {Hellier}, {Jehin}, {Lendl},
  {Maxted}, {Neveu-VanMalle}, {Pepe}, {Pollacco}, {Queloz}, {S{\'e}gransan},
  {Smalley}, {Smith}, {Triaud}, {Udry}, {Van Grootel}, \& {West}}]{Delrez2016}
{Delrez}, L., {Santerne}, A., {Almenara}, J.~M., {et~al.} 2016, \mnras, 458,
  4025, \dodoi{10.1093/mnras/stw522}

\bibitem[{{Dos Santos} {et~al.}(2023){Dos Santos}, {Alam}, {Espinoza}, \&
  {Vissapragada}}]{DosSantos2023}
{Dos Santos}, L.~A., {Alam}, M.~K., {Espinoza}, N., \& {Vissapragada}, S. 2023,
  \aj, 165, 244, \dodoi{10.3847/1538-3881/accf10}

\bibitem[{{Dos Santos} {et~al.}(2022){Dos Santos}, {Vidotto}, {Vissapragada},
  {Alam}, {Allart}, {Bourrier}, {Kirk}, {Seidel}, \&
  {Ehrenreich}}]{DSantos2022}
{Dos Santos}, L.~A., {Vidotto}, A.~A., {Vissapragada}, S., {et~al.} 2022, \aap,
  659, A62, \dodoi{10.1051/0004-6361/202142038}

\bibitem[{{Dragomir} {et~al.}(2020){Dragomir}, {Crossfield}, {Benneke}, {Wong},
  {Daylan}, {Diaz}, {Deming}, {Molliere}, {Kreidberg}, {Jenkins}, {Berardo},
  {Christiansen}, {Dressing}, {Gorjian}, {Kane}, {Mikal-Evans}, {Morales},
  {Werner}, {Ricker}, {Vanderspek}, {Seager}, {Winn}, {Jenkins}, {Col{\'o}n},
  {Fong}, {Guerrero}, {Hesse}, {Osborn}, {E. Rose}, {Smith}, \&
  {Ting}}]{Dragomir2020}
{Dragomir}, D., {Crossfield}, I. J.~M., {Benneke}, B., {et~al.} 2020, \apjl,
  903, L6, \dodoi{10.3847/2041-8213/abbc70}

\bibitem[{{Edwards} {et~al.}(2023){Edwards}, {Changeat}, {Tsiaras}, {Allan},
  {Behr}, {Hagey}, {Himes}, {Ma}, {Stassun}, {Thomas}, {Thompson}, {Boley},
  {Booth}, {Bouwman}, {France}, {Lowson}, {Meech}, {Phillips}, {Vidotto}, {Hou
  Yip}, {Bieger}, {Gressier}, {Janin}, {Jiang}, {Leonardi}, {Sarkar}, {Skaf},
  {Taylor}, {Yang}, \& {Ward-Thompson}}]{Edwards2023}
{Edwards}, B., {Changeat}, Q., {Tsiaras}, A., {et~al.} 2023, arXiv e-prints,
  arXiv:2306.13645, \dodoi{10.48550/arXiv.2306.13645}

\bibitem[{{Eggleton}(1983)}]{Eggleton1983}
{Eggleton}, P.~P. 1983, \apj, 268, 368, \dodoi{10.1086/160960}

\bibitem[{{Fern{\'a}ndez Fern{\'a}ndez} {et~al.}(2024){Fern{\'a}ndez
  Fern{\'a}ndez}, {Wheatley}, {King}, \& {Jenkins}}]{FernandezFernandez2024}
{Fern{\'a}ndez Fern{\'a}ndez}, J., {Wheatley}, P.~J., {King}, G.~W., \&
  {Jenkins}, J.~S. 2024, \mnras, 527, 911, \dodoi{10.1093/mnras/stad3263}

\bibitem[{{Foreman-Mackey} {et~al.}(2013){Foreman-Mackey}, {Hogg}, {Lang}, \&
  {Goodman}}]{Foreman-Mackey2013}
{Foreman-Mackey}, D., {Hogg}, D.~W., {Lang}, D., \& {Goodman}, J. 2013, \pasp,
  125, 306, \dodoi{10.1086/670067}

\bibitem[{{Fossati} {et~al.}(2023){Fossati}, {Pillitteri}, {Shaikhislamov},
  {Bonfanti}, {Borsa}, {Carleo}, {Guilluy}, \& {Rumenskikh}}]{Fossati2023}
{Fossati}, L., {Pillitteri}, I., {Shaikhislamov}, I.~F., {et~al.} 2023, \aap,
  673, A37, \dodoi{10.1051/0004-6361/202245667}

\bibitem[{{France} {et~al.}(2023){France}, {Fleming}, {Egan}, {Desert},
  {Fossati}, {Koskinen}, {Nell}, {Petit}, {Vidotto}, {Beasley}, {DeCicco},
  {Sreejith}, {Suresh}, {Baumert}, {Cauley}, {Villarreal D'Angelo}, {Hoadley},
  {Kane}, {Kohnert}, {Lambert}, \& {Ulrich}}]{France2023}
{France}, K., {Fleming}, B., {Egan}, A., {et~al.} 2023, \aj, 165, 63,
  \dodoi{10.3847/1538-3881/aca8a2}

\bibitem[{{Fu} {et~al.}(2022){Fu}, {Espinoza}, {Sing}, {Lothringer}, {Dos
  Santos}, {Rustamkulov}, {Deming}, {Kempton}, {Komacek}, {Knutson}, {Albert},
  {Pontoppidan}, {Volk}, \& {Filippazzo}}]{Fu2022}
{Fu}, G., {Espinoza}, N., {Sing}, D.~K., {et~al.} 2022, \apjl, 940, L35,
  \dodoi{10.3847/2041-8213/ac9977}

\bibitem[{{Gaia Collaboration} {et~al.}(2021){Gaia Collaboration}, {Brown},
  {Vallenari}, {Prusti}, {de Bruijne}, {Babusiaux}, {Biermann}, {Creevey},
  {Evans}, {Eyer}, \& et~al.}]{GaiaDR3}
{Gaia Collaboration}, {Brown}, A.~G.~A., {Vallenari}, A., {et~al.} 2021, \aap,
  649, A1, \dodoi{10.1051/0004-6361/202039657}

\bibitem[{{Gao} {et~al.}(2023){Gao}, {Piette}, {Steinrueck}, {Nixon}, {Zhang},
  {Kempton}, {Bean}, {Rauscher}, {Parmentier}, {Batalha}, {Savel}, {Arnold},
  {Roman}, {Malsky}, \& {Taylor}}]{Gao2023}
{Gao}, P., {Piette}, A. A.~A., {Steinrueck}, M.~E., {et~al.} 2023, \apj, 951,
  96, \dodoi{10.3847/1538-4357/acd16f}

\bibitem[{{Gordon} {et~al.}(2022){Gordon}, {Rothman}, {Hargreaves}, {Hashemi},
  {Karlovets}, {Skinner}, {Conway}, {Hill}, {Kochanov}, {Tan}, {Wcis{\l}o},
  {Finenko}, {Nelson}, {Bernath}, {Birk}, {Boudon}, {Campargue}, {Chance},
  {Coustenis}, {Drouin}, {Flaud}, {Gamache}, {Hodges}, {Jacquemart}, {Mlawer},
  {Nikitin}, {Perevalov}, {Rotger}, {Tennyson}, {Toon}, {Tran}, {Tyuterev},
  {Adkins}, {Baker}, {Barbe}, {Can{\`e}}, {Cs{\'a}sz{\'a}r}, {Dudaryonok},
  {Egorov}, {Fleisher}, {Fleurbaey}, {Foltynowicz}, {Furtenbacher}, {Harrison},
  {Hartmann}, {Horneman}, {Huang}, {Karman}, {Karns}, {Kassi}, {Kleiner},
  {Kofman}, {Kwabia-Tchana}, {Lavrentieva}, {Lee}, {Long}, {Lukashevskaya},
  {Lyulin}, {Makhnev}, {Matt}, {Massie}, {Melosso}, {Mikhailenko}, {Mondelain},
  {M{\"u}ller}, {Naumenko}, {Perrin}, {Polyansky}, {Raddaoui}, {Raston},
  {Reed}, {Rey}, {Richard}, {T{\'o}bi{\'a}s}, {Sadiek}, {Schwenke},
  {Starikova}, {Sung}, {Tamassia}, {Tashkun}, {Vander Auwera}, {Vasilenko},
  {Vigasin}, {Villanueva}, {Vispoel}, {Wagner}, {Yachmenev}, \&
  {Yurchenko}}]{Gordon2022}
{Gordon}, I.~E., {Rothman}, L.~S., {Hargreaves}, R.~J., {et~al.} 2022, \jqsrt,
  277, 107949, \dodoi{10.1016/j.jqsrt.2021.107949}

\bibitem[{{Grunblatt} {et~al.}(2023){Grunblatt}, {Saunders}, {Huber},
  {Thorngren}, {Vissapragada}, {Yoshida}, {Schlaufman}, {Giacalone},
  {MacDougall}, {Chontos}, {Turtelboom}, {Beard}, {Akana Murphy}, {Rice},
  {Isaacson}, {Angus}, \& {Howard}}]{Grunblatt2023}
{Grunblatt}, S., {Saunders}, N., {Huber}, D., {et~al.} 2023, arXiv e-prints,
  arXiv:2303.06728, \dodoi{10.48550/arXiv.2303.06728}

\bibitem[{{Guillochon} {et~al.}(2011){Guillochon}, {Ramirez-Ruiz}, \&
  {Lin}}]{Guillochon2011}
{Guillochon}, J., {Ramirez-Ruiz}, E., \& {Lin}, D. 2011, \apj, 732, 74,
  \dodoi{10.1088/0004-637X/732/2/74}

\bibitem[{{Guilluy} {et~al.}(2023){Guilluy}, {Bourrier}, {Jaziri}, {Dethier},
  {Mounzer}, {Giacobbe}, {Attia}, {Allart}, {Bonomo}, {Dos Santos}, {Rainer},
  \& {Sozzetti}}]{Guilluy2023}
{Guilluy}, G., {Bourrier}, V., {Jaziri}, Y., {et~al.} 2023, arXiv e-prints,
  arXiv:2307.00967, \dodoi{10.48550/arXiv.2307.00967}

\bibitem[{{Harris} {et~al.}(2020){Harris}, {Millman}, {van der Walt},
  {Gommers}, {Virtanen}, {Cournapeau}, {Wieser}, {Taylor}, {Berg}, {Smith},
  {Kern}, {Picus}, {Hoyer}, {van Kerkwijk}, {Brett}, {Haldane}, {del R{\'\i}o},
  {Wiebe}, {Peterson}, {G{\'e}rard-Marchant}, {Sheppard}, {Reddy}, {Weckesser},
  {Abbasi}, {Gohlke}, \& {Oliphant}}]{numpy}
{Harris}, C.~R., {Millman}, K.~J., {van der Walt}, S.~J., {et~al.} 2020, \nat,
  585, 357, \dodoi{10.1038/s41586-020-2649-2}

\bibitem[{{Higson} {et~al.}(2019){Higson}, {Handley}, {Hobson}, \&
  {Lasenby}}]{Higson2019}
{Higson}, E., {Handley}, W., {Hobson}, M., \& {Lasenby}, A. 2019, Statistics
  and Computing, 29, 891, \dodoi{10.1007/s11222-018-9844-0}

\bibitem[{{Hoyer} {et~al.}(2023){Hoyer}, {Jenkins}, {Parmentier}, {Deleuil},
  {Scandariato}, {Wilson}, {D{\'\i}az}, {Crossfield}, {Dragomir}, {Kataria},
  {Lendl}, {Ramirez}, {Pe{\~n}a Rojas}, \& {Vin{\'e}s}}]{Hoyer2023}
{Hoyer}, S., {Jenkins}, J.~S., {Parmentier}, V., {et~al.} 2023, \aap, 675, A81,
  \dodoi{10.1051/0004-6361/202346117}

\bibitem[{{Huang} {et~al.}(2023){Huang}, {Koskinen}, {Lavvas}, \&
  {Fossati}}]{Huang2023}
{Huang}, C., {Koskinen}, T., {Lavvas}, P., \& {Fossati}, L. 2023, arXiv
  e-prints, arXiv:2304.07352, \dodoi{10.48550/arXiv.2304.07352}

\bibitem[{Hunter(2007)}]{Hunter2007}
Hunter, J.~D. 2007, Computing in Science \& Engineering, 9, 90,
  \dodoi{10.1109/MCSE.2007.55}

\bibitem[{{Ikeda} {et~al.}(2016){Ikeda}, {Kobayashi}, {Kondo}, {Otsubo},
  {Hamano}, {Sameshima}, {Yoshikawa}, {Fukue}, {Nakanishi}, {Kawanishi},
  {Nakaoka}, {Kinoshita}, {Kitano}, {Asano}, {Takenaka}, {Watase}, {Mito},
  {Yasui}, {Minami}, {Izumu}, {Yamamoto}, {Mizumoto}, {Arasaki}, {Arai},
  {Matsunaga}, \& {Kawakita}}]{Ikeda2016}
{Ikeda}, Y., {Kobayashi}, N., {Kondo}, S., {et~al.} 2016, in Society of
  Photo-Optical Instrumentation Engineers (SPIE) Conference Series, Vol. 9908,
  Ground-based and Airborne Instrumentation for Astronomy VI, ed. C.~J.
  {Evans}, L.~{Simard}, \& H.~{Takami}, 99085Z, \dodoi{10.1117/12.2230886}

\bibitem[{{Ikeda} {et~al.}(2022){Ikeda}, {Kondo}, {Otsubo}, {Hamano}, {Yasui},
  {Matsunaga}, {Sameshima}, {Yoshikawa}, {Fukue}, {Nakanishi}, {Kawanishi},
  {Watase}, {Nakaoka}, {Arai}, {Kinoshita}, {Kitano}, {Nakamura}, {Asano},
  {Takenaka}, {Murai}, {Kawakita}, {Minami}, {Izumi}, {Yamamoto}, {Mizumoto},
  {Taniguchi}, \& {Tsujimoto}}]{Ikeda2022}
{Ikeda}, Y., {Kondo}, S., {Otsubo}, S., {et~al.} 2022, \pasp, 134, 015004,
  \dodoi{10.1088/1538-3873/ac1c5f}

\bibitem[{{Ionov} {et~al.}(2018){Ionov}, {Pavlyuchenkov}, \&
  {Shematovich}}]{Ionov2018}
{Ionov}, D.~E., {Pavlyuchenkov}, Y.~N., \& {Shematovich}, V.~I. 2018, \mnras,
  476, 5639, \dodoi{10.1093/mnras/sty626}

\bibitem[{{Ito} \& {Ikoma}(2021)}]{Ito2021}
{Ito}, Y., \& {Ikoma}, M. 2021, \mnras, 502, 750,
  \dodoi{10.1093/mnras/staa3962}

\bibitem[{{Jenkins} {et~al.}(2020){Jenkins}, {D{\'\i}az}, {Kurtovic},
  {Espinoza}, {Vines}, {Rojas}, {Brahm}, {Torres}, {Cort{\'e}s-Zuleta}, {Soto},
  {Lopez}, {King}, {Wheatley}, {Winn}, {Ciardi}, {Ricker}, {Vanderspek},
  {Latham}, {Seager}, {Jenkins}, {Beichman}, {Bieryla}, {Burke},
  {Christiansen}, {Henze}, {Klaus}, {McCauliff}, {Mori}, {Narita}, {Nishiumi},
  {Tamura}, {de Leon}, {Quinn}, {Villase{\~n}or}, {Vezie}, {Lissauer},
  {Collins}, {Collins}, {Isopi}, {Mallia}, {Ercolino}, {Petrovich},
  {Jord{\'a}n}, {Acton}, {Armstrong}, {Bayliss}, {Bouchy}, {Belardi}, {Bryant},
  {Burleigh}, {Cabrera}, {Casewell}, {Chaushev}, {Cooke}, {Eigm{\"u}ller},
  {Erikson}, {Foxell}, {G{\"a}nsicke}, {Gill}, {Gillen}, {G{\"u}nther}, {Goad},
  {Hooton}, {Jackman}, {Louden}, {McCormac}, {Moyano}, {Nielsen}, {Pollacco},
  {Queloz}, {Rauer}, {Raynard}, {Smith}, {Tilbrook}, {Titz-Weider}, {Turner},
  {Udry}, {Walker}, {Watson}, {West}, {Palle}, {Ziegler}, {Law}, \&
  {Mann}}]{Jenkins2020}
{Jenkins}, J.~S., {D{\'\i}az}, M.~R., {Kurtovic}, N.~T., {et~al.} 2020, Nature
  Astronomy, 4, 1148, \dodoi{10.1038/s41550-020-1142-z}

\bibitem[{{Jensen} {et~al.}(2012){Jensen}, {Redfield}, {Endl}, {Cochran},
  {Koesterke}, \& {Barman}}]{Jensen2012}
{Jensen}, A.~G., {Redfield}, S., {Endl}, M., {et~al.} 2012, \apj, 751, 86,
  \dodoi{10.1088/0004-637X/751/2/86}

\bibitem[{{Kasper} {et~al.}(2020){Kasper}, {Bean}, {Oklop{\v{c}}i{\'c}},
  {Malsky}, {Kempton}, {D{\'e}sert}, {Rogers}, \& {Mansfield}}]{Kasper2020}
{Kasper}, D., {Bean}, J.~L., {Oklop{\v{c}}i{\'c}}, A., {et~al.} 2020, \aj, 160,
  258, \dodoi{10.3847/1538-3881/abbee6}

\bibitem[{{Kempton} {et~al.}(2023){Kempton}, {Zhang}, {Bean}, {Steinrueck},
  {Piette}, {Parmentier}, {Malsky}, {Roman}, {Rauscher}, {Gao}, {Bell}, {Xue},
  {Taylor}, {Savel}, {Arnold}, {Nixon}, {Stevenson}, {Mansfield}, {Kendrew},
  {Zieba}, {Ducrot}, {Dyrek}, {Lagage}, {Stassun}, {Henry}, {Barman}, {Lupu},
  {Malik}, {Kataria}, {Ih}, {Fu}, {Welbanks}, \& {McGill}}]{Kempton2023}
{Kempton}, E. M.~R., {Zhang}, M., {Bean}, J.~L., {et~al.} 2023, \nat, 620, 67,
  \dodoi{10.1038/s41586-023-06159-5}

\bibitem[{{Kochanov} {et~al.}(2016){Kochanov}, {Gordon}, {Rothman},
  {Wcis{\l}o}, {Hill}, \& {Wilzewski}}]{Kochanov2016}
{Kochanov}, R.~V., {Gordon}, I.~E., {Rothman}, L.~S., {et~al.} 2016, \jqsrt,
  177, 15, \dodoi{10.1016/j.jqsrt.2016.03.005}

\bibitem[{{Koskinen} {et~al.}(2022){Koskinen}, {Lavvas}, {Huang}, {Bergsten},
  {Fernandes}, \& {Young}}]{Koskinen2022}
{Koskinen}, T.~T., {Lavvas}, P., {Huang}, C., {et~al.} 2022, \apj, 929, 52,
  \dodoi{10.3847/1538-4357/ac4f45}

\bibitem[{{Kubyshkina} {et~al.}(2018){Kubyshkina}, {Fossati}, {Erkaev},
  {Johnstone}, {Cubillos}, {Kislyakova}, {Lammer}, {Lendl}, \&
  {Odert}}]{Kubyshkina2018}
{Kubyshkina}, D., {Fossati}, L., {Erkaev}, N.~V., {et~al.} 2018, \aap, 619,
  A151, \dodoi{10.1051/0004-6361/201833737}

\bibitem[{{Kurokawa} \& {Nakamoto}(2014)}]{Kurokawa2014}
{Kurokawa}, H., \& {Nakamoto}, T. 2014, \apj, 783, 54,
  \dodoi{10.1088/0004-637X/783/1/54}

\bibitem[{{Lamp{\'o}n} {et~al.}(2020){Lamp{\'o}n}, {L{\'o}pez-Puertas}, {Lara},
  {S{\'a}nchez-L{\'o}pez}, {Salz}, {Czesla}, {Sanz-Forcada}, {Molaverdikhani},
  {Alonso-Floriano}, {Nortmann}, {Caballero}, {Bauer}, {Pall{\'e}}, {Montes},
  {Quirrenbach}, {Nagel}, {Ribas}, {Reiners}, \& {Amado}}]{Lampon2020}
{Lamp{\'o}n}, M., {L{\'o}pez-Puertas}, M., {Lara}, L.~M., {et~al.} 2020, \aap,
  636, A13, \dodoi{10.1051/0004-6361/201937175}

\bibitem[{{Lamp{\'o}n} {et~al.}(2023){Lamp{\'o}n}, {L{\'o}pez-Puertas},
  {Sanz-Forcada}, {Czesla}, {Nortmann}, {Casasayas-Barris}, {Orell-Miquel},
  {S{\'a}nchez-L{\'o}pez}, {Danielski}, {Pall{\'e}}, {Molaverdikhani},
  {Henning}, {Caballero}, {Amado}, {Quirrenbach}, {Reiners}, \&
  {Ribas}}]{Lampon2023}
{Lamp{\'o}n}, M., {L{\'o}pez-Puertas}, M., {Sanz-Forcada}, J., {et~al.} 2023,
  \aap, 673, A140, \dodoi{10.1051/0004-6361/202245649}

\bibitem[{{Li} {et~al.}(2010){Li}, {Miller}, {Lin}, \& {Fortney}}]{Li2010}
{Li}, S.-L., {Miller}, N., {Lin}, D. N.~C., \& {Fortney}, J.~J. 2010, \nat,
  463, 1054, \dodoi{10.1038/nature08715}

\bibitem[{{Linsky} {et~al.}(2014){Linsky}, {Fontenla}, \&
  {France}}]{Linsky2014}
{Linsky}, J.~L., {Fontenla}, J., \& {France}, K. 2014, \apj, 780, 61,
  \dodoi{10.1088/0004-637X/780/1/61}

\bibitem[{{Linssen} \& {Oklop{\v{c}}i{\'c}}(2023)}]{Linssen2023}
{Linssen}, D.~C., \& {Oklop{\v{c}}i{\'c}}, A. 2023, \aap, 675, A193,
  \dodoi{10.1051/0004-6361/202346583}

\bibitem[{{Lopez}(2017)}]{Lopez2017}
{Lopez}, E.~D. 2017, \mnras, 472, 245, \dodoi{10.1093/mnras/stx1558}

\bibitem[{{MacLeod} \& {Oklop{\v{c}}i{\'c}}(2022)}]{MacLeod2022}
{MacLeod}, M., \& {Oklop{\v{c}}i{\'c}}, A. 2022, \apj, 926, 226,
  \dodoi{10.3847/1538-4357/ac46ce}

\bibitem[{{Matsakos} \& {K{\"o}nigl}(2016)}]{Matsakos2016}
{Matsakos}, T., \& {K{\"o}nigl}, A. 2016, \apjl, 820, L8,
  \dodoi{10.3847/2041-8205/820/1/L8}

\bibitem[{{Mazeh} {et~al.}(2016){Mazeh}, {Holczer}, \& {Faigler}}]{Mazeh2016}
{Mazeh}, T., {Holczer}, T., \& {Faigler}, S. 2016, \aap, 589, A75,
  \dodoi{10.1051/0004-6361/201528065}

\bibitem[{{Nakayama} {et~al.}(2022){Nakayama}, {Ikoma}, \&
  {Terada}}]{Nakayama2022}
{Nakayama}, A., {Ikoma}, M., \& {Terada}, N. 2022, \apj, 937, 72,
  \dodoi{10.3847/1538-4357/ac86ca}

\bibitem[{{Naponiello} {et~al.}(2023){Naponiello}, {Mancini}, {Sozzetti},
  {Bonomo}, {Morbidelli}, {Dou}, {Zeng}, {Leinhardt}, {Biazzo}, {Cubillos},
  {Pinamonti}, {Locci}, {Maggio}, {Damasso}, {Lanza}, {Lissauer}, {Collins},
  {Carter}, {Jensen}, {Bignamini}, {Boschin}, {Bouma}, {Ciardi}, {Cosentino},
  {Crossfield}, {Desidera}, {Dumusque}, {Fiorenzano}, {Fukui}, {Giacobbe},
  {Gnilka}, {Ghedina}, {Guilluy}, {Harutyunyan}, {Howell}, {Jenkins}, {Lund},
  {Kielkopf}, {Lester}, {Malavolta}, {Mann}, {Matson}, {Matthews}, {Nardiello},
  {Narita}, {Pace}, {Pagano}, {Palle}, {Pedani}, {Seager}, {Schlieder},
  {Schwarz}, {Shporer}, {Twicken}, {Winn}, {Ziegler}, \&
  {Zingales}}]{Naponiello2023}
{Naponiello}, L., {Mancini}, L., {Sozzetti}, A., {et~al.} 2023, \nat, 622, 255,
  \dodoi{10.1038/s41586-023-06499-2}

\bibitem[{{Oklop{\v{c}}i{\'c}}(2019)}]{Oklopcic2019}
{Oklop{\v{c}}i{\'c}}, A. 2019, \apj, 881, 133, \dodoi{10.3847/1538-4357/ab2f7f}

\bibitem[{{Oklop{\v{c}}i{\'c}} \& {Hirata}(2018)}]{Oklopcic2018}
{Oklop{\v{c}}i{\'c}}, A., \& {Hirata}, C.~M. 2018, \apjl, 855, L11,
  \dodoi{10.3847/2041-8213/aaada9}

\bibitem[{{Orell-Miquel} {et~al.}(2022){Orell-Miquel}, {Murgas}, {Pall{\'e}},
  {Lamp{\'o}n}, {L{\'o}pez-Puertas}, {Sanz-Forcada}, {Nagel}, {Kaminski},
  {Casasayas-Barris}, {Nortmann}, {Luque}, {Molaverdikhani}, {Sedaghati},
  {Caballero}, {Amado}, {Bergond}, {Czesla}, {Hatzes}, {Henning},
  {Khalafinejad}, {Montes}, {Morello}, {Quirrenbach}, {Reiners}, {Ribas},
  {S{\'a}nchez-L{\'o}pez}, {Schweitzer}, {Stangret}, {Yan}, \& {Zapatero
  Osorio}}]{Orell-Miquel2022}
{Orell-Miquel}, J., {Murgas}, F., {Pall{\'e}}, E., {et~al.} 2022, \aap, 659,
  A55, \dodoi{10.1051/0004-6361/202142455}

\bibitem[{{Osborn} {et~al.}(2023){Osborn}, {Armstrong}, {Fern{\'a}ndez
  Fern{\'a}ndez}, {Knierim}, {Adibekyan}, {Collins}, {Delgado-Mena},
  {Fridlund}, {Gomes da Silva}, {Hellier}, {Jackson}, {King}, {Lillo-Box},
  {Matson}, {Matthews}, {Santos}, {Sousa}, {Stassun}, {Tan}, {Ricker},
  {Vanderspek}, {Latham}, {Seager}, {Winn}, {Jenkins}, {Bayliss}, {Bouma},
  {Ciardi}, {Collins}, {Col{\'o}n}, {Crossfield}, {Demangeon}, {D{\'\i}az},
  {Dorn}, {Dumusque}, {Aron Fetzner Keniger}, {Figueira}, {Gan}, {Goeke},
  {Hadjigeorghiou}, {Hawthorn}, {Helled}, {Howell}, {Nielsen}, {Osborn},
  {Quinn}, {Sefako}, {Shporer}, {Str{\o}m}, {Twicken}, {Vanderburg}, \&
  {Wheatley}}]{Osborn2023}
{Osborn}, A., {Armstrong}, D.~J., {Fern{\'a}ndez Fern{\'a}ndez}, J., {et~al.}
  2023, arXiv e-prints, arXiv:2308.12137, \dodoi{10.48550/arXiv.2308.12137}

\bibitem[{{Otsubo} {et~al.}(2016){Otsubo}, {Ikeda}, {Kobayashi}, {Sukegawa},
  {Kondo}, {Hamano}, {Sameshima}, {Fukue}, {Yoshikawa}, {Nakanishi}, {Watase},
  {Takenaka}, {Asano}, {Yasui}, {Matsunaga}, \& {Kawakita}}]{Otsubo2016}
{Otsubo}, S., {Ikeda}, Y., {Kobayashi}, N., {et~al.} 2016, in Society of
  Photo-Optical Instrumentation Engineers (SPIE) Conference Series, Vol. 9908,
  Ground-based and Airborne Instrumentation for Astronomy VI, ed. C.~J.
  {Evans}, L.~{Simard}, \& H.~{Takami}, 990879, \dodoi{10.1117/12.2233845}

\bibitem[{{Owen} \& {Jackson}(2012)}]{Owen2012}
{Owen}, J.~E., \& {Jackson}, A.~P. 2012, \mnras, 425, 2931,
  \dodoi{10.1111/j.1365-2966.2012.21481.x}

\bibitem[{{Owen} \& {Lai}(2018)}]{OwenLai2018}
{Owen}, J.~E., \& {Lai}, D. 2018, \mnras, 479, 5012,
  \dodoi{10.1093/mnras/sty1760}

\bibitem[{{Owen} \& {Murray-Clay}(2018)}]{Owen2018}
{Owen}, J.~E., \& {Murray-Clay}, R. 2018, \mnras, 480, 2206,
  \dodoi{10.1093/mnras/sty1943}

\bibitem[{{Persson} {et~al.}(2022){Persson}, {Georgieva}, {Gandolfi}, {Acuna},
  {Aguichine}, {Muresan}, {Guenther}, {Livingston}, {Collins}, {Dai},
  {Fridlund}, {Goffo}, {Jenkins}, {Kab{\'a}th}, {Korth}, {Levine}, {Serrano},
  {Vines}, {Barragan}, {Carleo}, {Colon}, {Cochran}, {Christiansen}, {Deeg},
  {Deleuil}, {Dragomir}, {Esposito}, {Gan}, {Grziwa}, {Hatzes}, {Hesse},
  {Horne}, {Jenkins}, {Kielkopf}, {Klagyivik}, {Lam}, {Latham}, {Luque},
  {Orell-Miquel}, {Mortier}, {Mousis}, {Narita}, {Osborne}, {Palle}, {Papini},
  {Ricker}, {Schmerling}, {Seager}, {Stassun}, {Van Eylen}, {Vanderspek},
  {Wang}, {Winn}, {Wohler}, {Zambelli}, \& {Ziegler}}]{Persson2022}
{Persson}, C.~M., {Georgieva}, I.~Y., {Gandolfi}, D., {et~al.} 2022, \aap, 666,
  A184, \dodoi{10.1051/0004-6361/202244118}

\bibitem[{{Pont} {et~al.}(2006){Pont}, {Zucker}, \& {Queloz}}]{Pont2006}
{Pont}, F., {Zucker}, S., \& {Queloz}, D. 2006, \mnras, 373, 231,
  \dodoi{10.1111/j.1365-2966.2006.11012.x}

\bibitem[{{Rappaport} {et~al.}(2013){Rappaport}, {Sanchis-Ojeda}, {Rogers},
  {Levine}, \& {Winn}}]{Rappaport2013}
{Rappaport}, S., {Sanchis-Ojeda}, R., {Rogers}, L.~A., {Levine}, A., \& {Winn},
  J.~N. 2013, \apjl, 773, L15, \dodoi{10.1088/2041-8205/773/1/L15}

\bibitem[{{Salz} {et~al.}(2016){Salz}, {Schneider}, {Czesla}, \&
  {Schmitt}}]{Salz2016}
{Salz}, M., {Schneider}, P.~C., {Czesla}, S., \& {Schmitt}, J.~H.~M.~M. 2016,
  \aap, 585, L2, \dodoi{10.1051/0004-6361/201527042}

\bibitem[{{Schreyer} {et~al.}(2023){Schreyer}, {Owen}, {Spake}, {Bahroloom}, \&
  {Di Giampasquale}}]{Schreyer2023}
{Schreyer}, E., {Owen}, J.~E., {Spake}, J.~J., {Bahroloom}, Z., \& {Di
  Giampasquale}, S. 2023, arXiv e-prints, arXiv:2302.10947,
  \dodoi{10.48550/arXiv.2302.10947}

\bibitem[{{Sing} {et~al.}(2019){Sing}, {Lavvas}, {Ballester}, {Lecavelier des
  Etangs}, {Marley}, {Nikolov}, {Ben-Jaffel}, {Bourrier}, {Buchhave}, {Deming},
  {Ehrenreich}, {Mikal-Evans}, {Kataria}, {Lewis}, {L{\'o}pez-Morales},
  {Garc{\'\i}a Mu{\~n}oz}, {Henry}, {Sanz-Forcada}, {Spake}, {Wakeford}, \&
  {PanCET Collaboration}}]{Sing2019}
{Sing}, D.~K., {Lavvas}, P., {Ballester}, G.~E., {et~al.} 2019, \aj, 158, 91,
  \dodoi{10.3847/1538-3881/ab2986}

\bibitem[{{Skilling}(2004)}]{Skilling2004}
{Skilling}, J. 2004, in American Institute of Physics Conference Series, Vol.
  735, Bayesian Inference and Maximum Entropy Methods in Science and
  Engineering: 24th International Workshop on Bayesian Inference and Maximum
  Entropy Methods in Science and Engineering, ed. R.~{Fischer}, R.~{Preuss}, \&
  U.~V. {Toussaint}, 395--405, \dodoi{10.1063/1.1835238}

\bibitem[{Skilling(2006)}]{Skilling2006}
Skilling, J. 2006, Bayesian Analysis, 1, 833 , \dodoi{10.1214/06-BA127}

\bibitem[{{Spake} {et~al.}(2021){Spake}, {Oklop{\v{c}}i{\'c}}, \&
  {Hillenbrand}}]{Spake2021}
{Spake}, J.~J., {Oklop{\v{c}}i{\'c}}, A., \& {Hillenbrand}, L.~A. 2021, \aj,
  162, 284, \dodoi{10.3847/1538-3881/ac178a}

\bibitem[{{Spake} {et~al.}(2018){Spake}, {Sing}, {Evans}, {Oklop{\v{c}}i{\'c}},
  {}, {Bourrier}, {Kreidberg}, {Rackham}, {Irwin}, {Ehrenreich}, {Wyttenbach},
  {Wakeford}, {Zhou}, {Chubb}, {Nikolov}, {Goyal}, {Henry}, {Williamson},
  {Blumenthal}, {Anderson}, {Hellier}, {Charbonneau}, {Udry}, \&
  {Madhusudhan}}]{Spake2018}
{Spake}, J.~J., {Sing}, D.~K., {Evans}, T.~M., {et~al.} 2018, \nat, 557, 68,
  \dodoi{10.1038/s41586-018-0067-5}

\bibitem[{{Spake} {et~al.}(2022){Spake}, {Oklop{\v{c}}i{\'c}}, {Hillenbrand},
  {Knutson}, {Kasper}, {Dai}, {Orell-Miquel}, {Vissapragada}, {Zhang}, \&
  {Bean}}]{Spake2022}
{Spake}, J.~J., {Oklop{\v{c}}i{\'c}}, A., {Hillenbrand}, L.~A., {et~al.} 2022,
  \apjl, 939, L11, \dodoi{10.3847/2041-8213/ac88c9}

\bibitem[{{Speagle}(2020)}]{Speagle2020}
{Speagle}, J.~S. 2020, \mnras, 493, 3132, \dodoi{10.1093/mnras/staa278}

\bibitem[{{Sreejith} {et~al.}(2023){Sreejith}, {France}, {Fossati}, {Koskinen},
  {Egan}, {Cauley}, {Cubillos}, {Ambily}, {Huang}, {Lavvas}, {Fleming},
  {Desert}, {Nell}, {Petit}, \& {Vidotto}}]{Sreejith2023}
{Sreejith}, A.~G., {France}, K., {Fossati}, L., {et~al.} 2023, \apjl, 954, L23,
  \dodoi{10.3847/2041-8213/acef1c}

\bibitem[{{Thorngren} {et~al.}(2023){Thorngren}, {Lee}, \&
  {Lopez}}]{Thorngren2023}
{Thorngren}, D.~P., {Lee}, E.~J., \& {Lopez}, E.~D. 2023, \apjl, 945, L36,
  \dodoi{10.3847/2041-8213/acbd35}

\bibitem[{Virtanen {et~al.}(2020)Virtanen, Gommers, Oliphant, Haberland, Reddy,
  Cournapeau, Burovski, Peterson, Weckesser, Bright, {van der Walt}, Brett,
  Wilson, Millman, Mayorov, Nelson, Jones, Kern, Larson, Carey, Polat, Feng,
  Moore, {VanderPlas}, Laxalde, Perktold, Cimrman, Henriksen, Quintero, Harris,
  Archibald, Ribeiro, Pedregosa, {van Mulbregt}, \& {SciPy 1.0
  Contributors}}]{scipy}
Virtanen, P., Gommers, R., Oliphant, T.~E., {et~al.} 2020, Nature Methods, 17,
  261, \dodoi{10.1038/s41592-019-0686-2}

\bibitem[{{Vissapragada} {et~al.}(2022){Vissapragada}, {Knutson},
  {Greklek-McKeon}, {Oklop{\v{c}}i{\'c}}, {Dai}, {dos Santos}, {Jovanovic},
  {Mawet}, {Millar-Blanchaer}, {Paragas}, {Spake}, {Tinyanont}, \&
  {Vasisht}}]{Vissapragada2022:helium}
{Vissapragada}, S., {Knutson}, H.~A., {Greklek-McKeon}, M., {et~al.} 2022, \aj,
  164, 234, \dodoi{10.3847/1538-3881/ac92f2}

\bibitem[{{Wang} \& {Dai}(2021)}]{Wang2021}
{Wang}, L., \& {Dai}, F. 2021, \apj, 914, 99, \dodoi{10.3847/1538-4357/abf1ed}

\bibitem[{{Wyttenbach} {et~al.}(2020){Wyttenbach}, {Molli{\`e}re},
  {Ehrenreich}, {Cegla}, {Bourrier}, {Lovis}, {Pino}, {Allart}, {Seidel},
  {Hoeijmakers}, {Nielsen}, {Lavie}, {Pepe}, {Bonfils}, \&
  {Snellen}}]{Wyttenbach2020}
{Wyttenbach}, A., {Molli{\`e}re}, P., {Ehrenreich}, D., {et~al.} 2020, \aap,
  638, A87, \dodoi{10.1051/0004-6361/201937316}

\bibitem[{{Yan} {et~al.}(2022){Yan}, {Seon}, {Guo}, {Chen}, \& {Li}}]{Yan2022}
{Yan}, D., {Seon}, K.-i., {Guo}, J., {Chen}, G., \& {Li}, L. 2022, \apj, 936,
  177, \dodoi{10.3847/1538-4357/ac8793}

\bibitem[{{Yan} \& {Henning}(2018)}]{Yan2018}
{Yan}, F., \& {Henning}, T. 2018, Nature Astronomy, 2, 714,
  \dodoi{10.1038/s41550-018-0503-3}

\bibitem[{{Youngblood} {et~al.}(2022){Youngblood}, {Pineda}, {Ayres}, {France},
  {Linsky}, {Wood}, {Redfield}, \& {Schlieder}}]{Youngblood2022}
{Youngblood}, A., {Pineda}, J.~S., {Ayres}, T., {et~al.} 2022, \apj, 926, 129,
  \dodoi{10.3847/1538-4357/ac4711}

\bibitem[{{Zhang} {et~al.}(2023){Zhang}, {Dai}, {Bean}, {Knutson}, \&
  {Rescigno}}]{Zhang2023}
{Zhang}, M., {Dai}, F., {Bean}, J.~L., {Knutson}, H.~A., \& {Rescigno}, F.
  2023, \apjl, 953, L25, \dodoi{10.3847/2041-8213/aced51}

\bibitem[{{Zhang} {et~al.}(2022){Zhang}, {Knutson}, {Wang}, {Dai}, \&
  {Barrag{\'a}n}}]{Zhang2022}
{Zhang}, M., {Knutson}, H.~A., {Wang}, L., {Dai}, F., \& {Barrag{\'a}n}, O.
  2022, \aj, 163, 67, \dodoi{10.3847/1538-3881/ac3fa7}

\bibitem[{{Zhang} {et~al.}(2021){Zhang}, {Knutson}, {Wang}, {Dai}, {Oklopcic},
  \& {Hu}}]{Zhang2021}
{Zhang}, M., {Knutson}, H.~A., {Wang}, L., {et~al.} 2021, \aj, 161, 181,
  \dodoi{10.3847/1538-3881/abe382}

\end{thebibliography}

\end{document}